\documentclass[aps,prl,twocolumn]{revtex4-1}
\usepackage{graphics,color,graphicx,amsmath}
\usepackage{xr}

\begin{document}

\newcommand{\mr}[1]{\mathrm{#1}}
\newcommand{\mb}[1]{\mathbf{#1}}
\newcommand{\br}[1]{\left<#1\right>}
\newcommand{\bl}[1]{\left|#1\right|}
\newcommand{\mc}[1]{\mathcal{#1}}
\newcommand{\tb}[1]{\textcolor{blue}{#1}}
\newcommand{\tr}[1]{\textcolor{red}{#1}}
\newcommand{\tg}[1]{\textcolor{green}{#1}}
\newcommand{\si}[1]{\sigma_{\rm #1}}

\title{Strong consensus on US Supreme Court spans a century}
\author{Edward D. Lee}
\affiliation{Department of Physics, 142 Science Dr, Cornell University, Ithaca NY 14850}

\date{\today}

\begin{abstract}
The US Supreme Court throughout the 20th century has been characterized as being divided between  liberals and conservatives, suggesting that justices with similar ideologies would have voted similarly had they overlapped in tenure. What if they had? I build an empirical, quantitative model of this counterfactual hypothesis using pairwise maximum entropy. I infer how 36 justices from 1946-2016 would have all voted on a Super Supreme Court. The model is strikingly consistent with a standard voting model from political science despite using $10^5$ less parameters and fitting the observed statistics better. As with historical courts, the Super Court is dominated by consensus. The rate at which consensus decays as more justices are included is extremely slow, nearly 100 years, and indicates that the modern Supreme Court is an extremely stable institution. Beyond consensus, I discover a rich structure of dissenting blocs that are distributed along a heavy-tailed Zipf's law. The heavy tail means that dominant dissenting modes fail to capture the entire spectrum of dissent. Thus, I find that Supreme Court voting over time is not low-dimensional despite implications to the contrary in historical analysis of Supreme Court voting. Although it has been long presumed that strong higher order correlations are induced by features of the cases, the institution, and the justices, I show that such complexity can be expressed in a minimal model relying only on pairwise correlations. From the perspective of model selection, this minimal model may generalize better and thus be useful for prediction of Supreme Court voting over time.
\end{abstract}

\maketitle
The modern US Supreme Court is often described as being divided between liberal and conservative wings balanced on the fulcrum of one or two swing voters \cite{Brazill:2002te,Martin:2004vd,Lawson:2006hp}. In the 1930s, a similar dynamic was at work where a conservative bloc known as the ``Four Horsemen'' relied on a fifth swing vote to countermand policy designed to ameliorate the economic impact of the Great Depression \cite{Urofsky:2017ug}. Over time, political issues and the composition of the Court have changed, but the idea that a left and right divide characterizes voting on the Court is a widespread observation in the literature \cite{Sirovich:2003wt,Martin:2004vd,Tenenbaum:2008tn,Segal:2015dn,Lee:2015ev}. 
The implicit hypothesis is that ideologically similar justices on different courts would have voted similarly on the same cases, but in reality they never faced the same cases nor voted with the same set of colleagues. Although I could take similar cases and then compare two justices' votes \cite{Tate:1981wy,Segal:2015dn}, a direct approach would be to compare how the two vote with a shared cohort to infer how the two might have voted together.

The Supreme Court is the highest court in the US, consisting of up to 9 sitting justices at any given time with short periods where a seat is vacant. Each set of sitting justices is known as a \textit{natural court}. Unlike Congress where all seats come up for reelection every few years, Supreme Court justices are appointed for life and typically one justice is replaced at a time. Although a retiring and an incoming justice may never have voted together, they may both accumulate a number of votes with the same 8 justices over many years and potentially many more years with a smaller cohort. In a typical year, 90-150 cases are decided with a vote (depending on the era) \cite{SupremeCourtD:YPR-mEko,States:TdVwx51P}. After choosing to consider a case, the Opinion of the Court is decided by a majority and---although there may be subtleties in the Opinion---the voting justices must in the end choose to join the majority or not, and this binary code is used in the political science literature \cite{Quinn:2017ji}. 

The long overlapping stints are informative about how justices vote relative to each other. If the two pairs out of three voters were highly correlated, then I would by transitivity expect that the third pair likewise vote together. If, however, the two pairs were anticorrelated, then the proverbial logic ``the enemy of my enemy is my friend'' implies that the third pair is again correlated. In general, there are an infinite number of models that would match the observed pairs while filling in the missing ones. If I insist that the model match the observed pairwise disagreements but otherwise make no further assumptions, I have specified pairwise maximum entropy (maxent) models. 
Maxent models by design make minimal assumptions about the structure of the data and so are not guaranteed to capture the statistics of political voting with few parameters. I show, however, that they do while remaining consistent with highly parameterized spatial voting models from political science \cite{Poole:1985fo}.

I build a pairwise maxent model of the joint voting record of a Super Supreme Court of all 36 justices in the modern Supreme Court Database from 1946-2016 ($K=8,737$) \cite{SupremeCourtD:YPR-mEko,Lee:2015ev}. Instead of ideological divisions, the Super Court is dominated by unanimity even across temporally separated justices, showing it to be an extremely stable institution. Just from pairwise correlations, the model generates a rich structure of dissenting blocs that cannot be captured by a low-dimensional, ideological picture.
This comparison of Supreme Court justices across time is just one example of the comparative study of political institutions. I show how minimal models from statistical physics make quantitative predictions that could be rigorously tested against alternative models or new data.

\begin{figure*}[tbp]\centering
	\includegraphics[width=\linewidth]{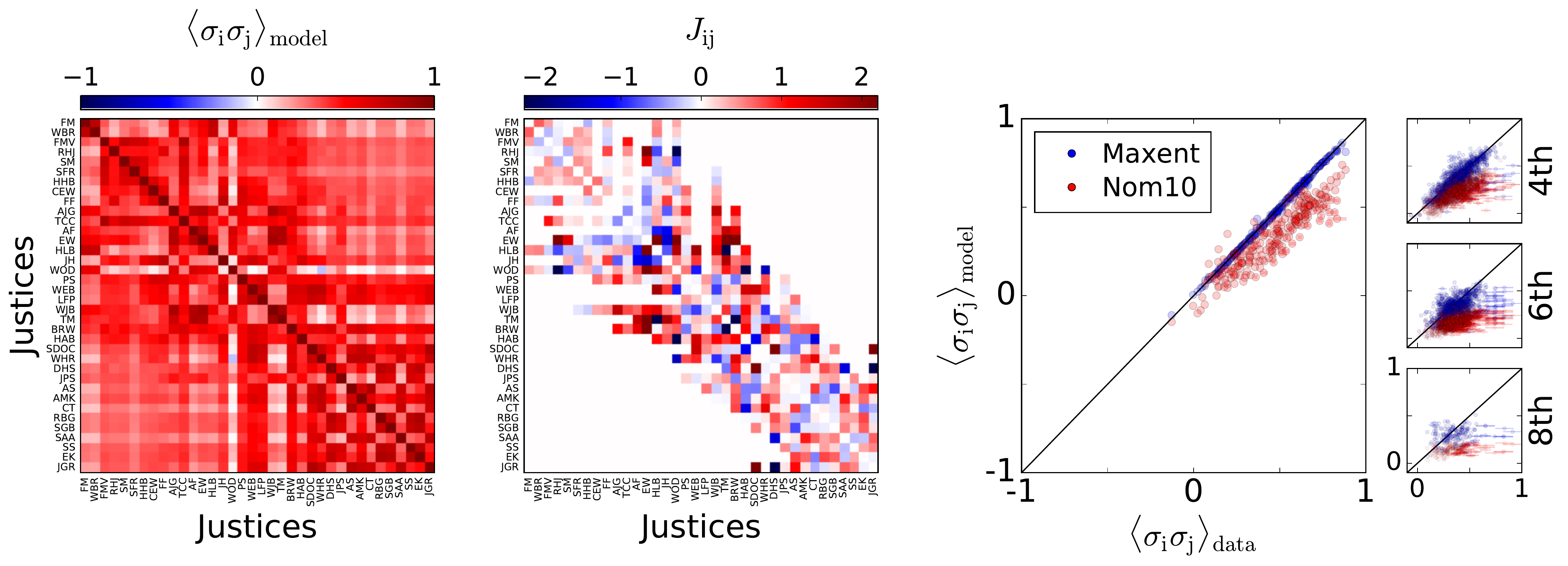}
\caption{(left) Predicted pairwise correlations from pairwise maxent model are almost all positive. (middle) Inferred couplings $J_{\rm ij}$. Pairs without any votes together in the data have $J_{\rm ij}=0$.
(right) Maxent model fits observed pairwise correlations well compared to the the 10-dimensional W-Nominate model after adjustment for unanimous votes specified by Eqns~\ref{eq:p unan 1} and \ref{eq:p unan 2}. For the maxent model, over 97\% of the 197 fit correlations are within two standard errors of the mean given \mbox{$\rm K_{\rm ij}$ observations for justices i and j, $\delta_{\rm ij} = \sqrt{(\br{\si{i}\si{j}}+1)(1-\br{\si{i}\si{j}})/4{\rm K}_{\rm ij}}$}, except for a few major deviations accounted for by conflicting constraints (Appendix Section~II). Maxent model predicts 4th, 6th, and 8th order correlations closely except some strong deviations explained by short temporal fluctuations in the distribution of case types (Appendix Section~III). For clarity, only points with absolute error \mbox{$\delta_{\rm ij...k} = \left|2\sqrt{p(\si{i}=\si{j}=...=\si{k})[1-p(\si{i}=\si{j}=...=\si{k})]/{\rm K}_{\rm ij...k}}\right|>0.25$} have error bars.}
\label{gr:sisj}
\end{figure*}

\section*{The maxent model}
I represent the vote of Justice i to be $\sigma_{\rm i}$ where the binary nature of majority voting is represented by \mbox{$\sigma_{\rm i}\in\{-1,1\}$}. The average agreement between two justices who voted $\rm K_{\rm ij}$ times together is measured by the pairwise correlation
\begin{align}
	\br{\sigma_{\rm i}\sigma_{\rm j}}_{\rm data} &= \sum_{\rm n=1}^{\rm K_{\rm ij}} \left.\sigma_{\rm i}^{\rm(n)}\sigma_{\rm j}^{\rm(n)}\right/\rm K_{\rm ij}.
\end{align}
I specify that the model match the observed pairwise correlations,
\begin{align}
	\br{\si{i}\si{j}}_{\rm data} &= \br{\si{i}\si{j}}= \sum_\sigma p(\sigma)\si{i}\si{j}\label{eq:pairs}
\end{align}
where the model is the probability distribution $p(\sigma)$ over a vector of votes $\sigma$ of all included justices. I must also specify the average vote of each justice $\br{\si{i}}$. For nearly all votes, the sign of the vote corresponds to whether to affirm or to reverse the previous court's ruling, but the orientation is determined by the history of the case. Since I focus on the pattern of internal agreement and do not expect that to depend on whether the question is formulated as ``Is A constitutional?'' or as ``Is A unconstitutional?'', I remove any bias by symmetrizing the data set such that \cite{Lee:2015ev,Quinn:2017ji}
\begin{align}
	\br{\si{i}}_{\rm data} &= \br{\si{i}}= 0 \label{eq:means}
\end{align}
Eqns \ref{eq:pairs} and \ref{eq:means} along with the principle of maximum entropy specify the values of the remaining unobserved pairwise correlations.

Entropy is the unique measure of uncertainty that obeys simple rules for consistency \cite{Shannon:1948wk,Cover:2006tl}. Maximizing entropy \mbox{$S[p]=-\sum_\sigma p(\sigma)\ln p(\sigma)$} while constraining the distribution to match the pairwise correlations is equivalent to minimizing the Helmholtz free energy \cite{Jaynes:1957fy,Bialek:2012ueb,Lee:2015ev}
\begin{align}
	-\ln Z &= \br{E}-S\\
	E(\sigma) &= -\sum_{\rm i<j}^{36} J_{\rm ij}\si{i}\si{j}\label{eq:E}
\end{align}
where the Hamiltonian is formally equivalent to the Ising spin glass and the couplings $J_{\rm ij}$ can be thought of as Langrangian multipliers enforcing Eq~\ref{eq:pairs}. The resulting pairwise maximum entropy model has Boltzmann form
\begin{align}
	p(\sigma) &= \left.e^{-E(\sigma)}\right/Z.
\end{align}
Despite this equivalence, the maxent formalism requires no assumptions about equilibrium, only that the pairwise correlations are representative of some stationary distribution $p(\sigma)$ I wish to characterize. 

For each $\br{\si{i}\si{j}}$ I fix, I have a corresponding coupling $J_{\rm ij}$ that, in the strictest sense, only has probabilistic implications. A positive $J_{\rm ij}$ lowers the energy $E$ when the two justices i and j vote together, increasing the probability that I observe such a vote by a factor $e^{J_{\rm ij}}$. Disagreement between i and j is suppressed by a factor $e^{-J_{\rm ij}}$. Since justices who never voted together have no correlations to constrain, the corresponding coupling are not specified, effectively setting $J_{\rm ij}=0$. Although one might interpret this to indicate the absence of real interaction, the precise statement is that the predicted correlations are mediated through the structure implied by the constrained correlations. The couplings do not imply causal interaction between the justices. If there were causal interactions, then the consequences would be hidden in the pairwise correlations along with any other effects like sampling biases. In other words, the only information in the model is that available in the pairwise correlations and no more as is explicitly imposed by the maximum entropy approach. 

In the maxent model for the Super Court, the pairwise correlations are almost exclusively positive, confirming that the Supreme Court has always been dominated by consensus. Out of the 248 observed pairwise correlations, only 2 are negative---between WO Douglas and WE Burger and between WO Douglas and WH Rehnquist---hinting that WO Douglas is an unusual justice in the history of the Court. After solving the model I can predict the hypothetical correlations between justices who never served together, and even in this enlarged set of possibilities the only further negative correlations are between WO Douglas and conservatives SD O'Connor, W Rehnquist, A Scalia, and C Thomas. Additionally, the ideologically opposed pairs WJ Brennan and C Thomas, T Marshall and WH Rehnquist, and T Marshall and C Thomas are negatively correlated. Although these are all small negative values, they are nevertheless prominent because they are unusual in a political body that prizes public agreement \cite{Epstein:2010wt,Urofsky:2017ug}. 
Although the remaining correlations are positive, the left-right divide is indicated by stronger correlations within and weaker correlations between ideological sides when the justices are ordered by ideology (Appendix Figure~4). This tension between competing blocs is reflected in the interaction matrix $J_{\rm ij}$, where couplings of both signs are recovered from our model of majority vs. minority voting.

\section{W-Nominate model}
I compare the maxent model with a standard spatial voting model from political science W-nominate (Appendix Section IV) \cite{Poole:1985fo}. Canonical spatial voting models assume that each justice is located in some low-dimensional, policy space along with the positions for every case considered. 
In this model, each justice votes independently, \mbox{$\log p(\sigma) = \sum_{\rm i}\log p(\si{i})\propto -\sum_{\rm i}\beta E(\sigma_{\rm i})$} with inverse temperature parameter $\beta$, and correlations are induced by similar positions in policy space.
Given justice i's preferred policy position $\theta_{\rm id}$ with distance from case k, $f_{\rm kd}(\theta_{\rm i})$, along dimension d of a D-dimensional space, every justice contributes to the total energy
\begin{align}
	E_{\rm i}(\sigma_{\rm i}) &= -\exp\left( -\sum_{\rm d=1}^{\rm D} {\rm w}_{\rm d}^2 f_{\rm kd}(\sigma_{\rm i})^2/2 \right)\label{eq:nom1}
\end{align}
The parameters $\rm w_{\rm d}$ weight each dimension. Eq~\ref{eq:nom1} stipulates that when cases are located close to a justice's preferred policy position, justices are more likely to vote for it than against it. 

The standard W-Nominate model excludes votes in the data where all justices voted together when fitting, so I include the probability of Super Court unanimity given by the pairwise maxent model $p_{\rm unan}$ by reweighting the W-Nominate model such that a unanimous vote $\sigma_{\rm u}$
\begin{align}
	p(\sigma_{\rm u}) &\leftarrow p(\sigma_{\rm u})(1-p_{\rm unan}) + p_{\rm unan}\label{eq:p unan 1}
\intertext{while other votes}
	p(\sigma) &\leftarrow p(\sigma)(1-p_{\rm unan}).\label{eq:p unan 2}
\end{align}
Eq~\ref{eq:p unan 1} does not fix the two models with the same $p(\sigma_{\rm u})$ because unanimous votes are generated with small probability by W-Nominate even though they are not fit (Appendix Section~IV), but it does ensure that if the W-Nominate model were like the maxent model {\it beyond} the unanimous mode, the similarities would be readily apparent.

The \mbox{$\rm D=10$} spatial voting model referred to as Nom10 has $>10^5$ parameters compared to \mbox{$<10^3$} for the pairwise maxent model even though it is minimal compared to other variations of spatial voting models. Despite this large disparity in parameterization, the maxent model fits the observed statistics of the data better as shown in Figure~\ref{gr:sisj}. Nevertheless, both models show very similar collective patterns.

\section*{Strong consensus}
\begin{figure}[b]\centering
	\includegraphics[width=\linewidth]{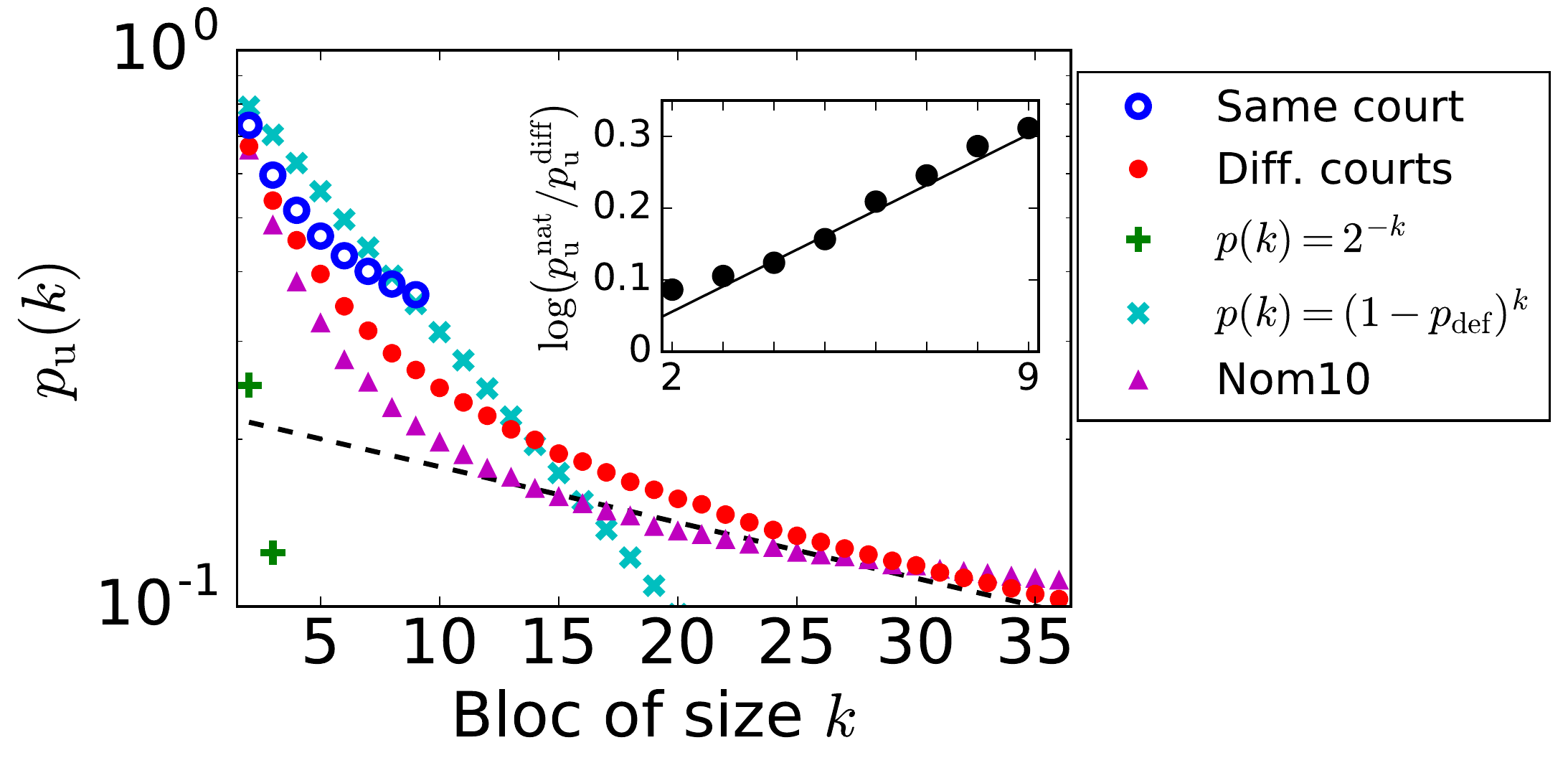}
\caption{Decay length of consensus spans a century in the hypothetical Super Court according to both maxent and Nom10 models. For the maxent model, the probability that all members of a bloc of size $k$ vote together $p_{\rm u}(k)$ decays exponentially when $k>20$. A fit to these points has a decay length of 43 justices (dashed black line) corresponding to nearly a century of justices.
I consider justices that sat on the same court (blue), justices who did not (red), independent voters (green), independent defectors from unanimity (cyan), and the 10-dimensional W-Nominate model (magenta).
(inset) ``Norm to consensus'' energy term pushing individuals to unanimity grows linearly $\log(p_{\rm u}^{\rm nat}/p_{\rm u}^{\rm diff}) \propto 0.035k$ for justices in the same natural court compared to justices who did not all work together. Consensus energy outweighs the tendency for unrelated justices to vote independently measured from the exponential tail that grows as $k/l = -0.023k$.
}
\label{gr:unanimity}
\end{figure}

Despite consisting of members from 24 nine-member natural courts, the probability of unanimity on the Super Court is an astounding 10\% of the time.
In comparison, independent justices would vote unanimously with a relatively minuscule probability of $p_{\rm u}(k=36) = 2^{-36}$, implying that the interactions impose strong collective structure. 
As a stricter test of the strength of cohesion, I might consider that justices already have an idea of how the Court is leaning before they cast their vote. If I instead imagine that each justice independently defects from unanimity with probability $p_{\rm def}$, I can measure for a natural court of $k'=9$ that typically votes unanimously about 36\% of the time, $p_{\rm def}=0.89$. Extrapolating this I find that the Super Court would vote unanimously with probability $p_{\rm u}(k=36) = 0.015$ \footnote{I use the probability of unanimity of a group of size $k'$ to estimate the probability that a single justice defects $p_{\rm def} = 1-p_{\rm u}^{1/k'}(k')$. The probability of no defections with $k$ independently defecting justices is $p_{\rm u}(k) = (1-p_{\rm def})^k$ as in Figure~\ref{gr:unanimity}.}. This is an order of magnitude below what I find on the Super Court. Thus, the Super Court shows a strong tendency to consensus that spans the many natural courts it represents.

Even in a chain of tightly correlated voters the probability that justices agree with others far away becomes dominated by drift \cite{Ising:1924vf}.
Looking at Figure~\ref{gr:sisj}, I see that pairwise correlations far from the diagonal tend to be smaller than those closer, indicative of weaker correlations spanning longer periods of time. 
To test this more systematically, I consider the probability of unanimity in blocs of size $k$, $p_{\rm u}(k)$, and compare justices that sat together with random subsets in Figure~\ref{gr:unanimity}. It is clear that both models of independent justices decay much more rapidly than what I find in the Super Court.
In the limit of large blocs of 20 or more justices, or roughly two different Supreme Courts, the unanimous mode begins to decay like an exponential, indicating the regime of finite correlation length. Here, adding a new justice is like adding an independent voice to the system. The decay length $l$, however, is 43 justices, roughly equivalent to a century of justices. W-Nominate predicts a similar decay length of 77 justices. Although the W-Nominate curve is lifted because I included unanimity in Eq~\ref{eq:p unan 1}, its shape which is not given is remarkably similar. In this quantitative sense, the Supreme Court is an extremely stable and conservative institution.

The force quelling dissent, a ``norm to consensus'' \cite{Epstein:2010wt}, is even stronger for justices that sit on the same court. I again calculate the probability of unanimity for blocks of size $k$, but only for justices who sat on the same natural court.
As $k$ increases from 2 to 9, the log-ratio of the probability, or energy difference, of finding a unanimous vote on a natural court $p_{\rm u}^{\rm nat}$ compared to justices from different courts $p_{\rm u}^{\rm diff}$ increases linearly with the bloc size $\log\left(p_{\rm u}^{\rm nat}(k)/p_{\rm u}^{\rm diff}(k)\right)=E^{\rm diff}(k)-E^{\rm nat}(k)\propto 0.035 k$ as I show in Figure~\ref{gr:unanimity}. This reveals a force encouraging the formation of blocs and especially for consensus specific to natural courts. Notably, the energy term for consensus grows faster than the inverse decorrelation length in the limit of large courts $l^{-1}=-0.023$ per justice. This suggests that consensus would dominate even when random justices who had never voted together are made to vote together which is akin to the situation when new justices are nominated into the Supreme Court. 

\section*{Zipf's law: Many ways to dissent}
Besides consensus, the pattern of dissenting blocs that emerge from pairwise interactions are consistent with notions of ideological blocs, but show much richer structure than binary ideological division. Ordering votes by frequency rank $r$, I find that Zipf's law holds after the strong unanimous mode \cite{Newman:2005gv,Schwab:2014io}
\begin{align}
	p(r) = (1-\alpha) r^{-\alpha}/(r_{\rm min}-r_{\rm max})\label{eq:zipf}
\end{align} 
where \mbox{$\alpha=0.85$}. 
The power law form with is clear over 4 orders of magnitude. This is consistent with previous work on the Second Rehnquist Court, where we observed a wide distribution of votes as measured from the data over a limited range of $\sim10^2$ unique votes. There, a power law fit yields $\alpha\approx 1$ \cite{Lee:2015ev}.

The slow decay in the distribution with $r$ signals that there is no typical set of voting behavior. The Super Court shows an extremely heavy-tailed distribution over votes where even ``rare'' events happen relatively often. In this sense, the intuitive notion that I could consider an average ideological behavior of the Super Court is misleading because even votes that defy this intuition are probable.

\begin{figure}[tbp]\centering
	\includegraphics[width=\linewidth]{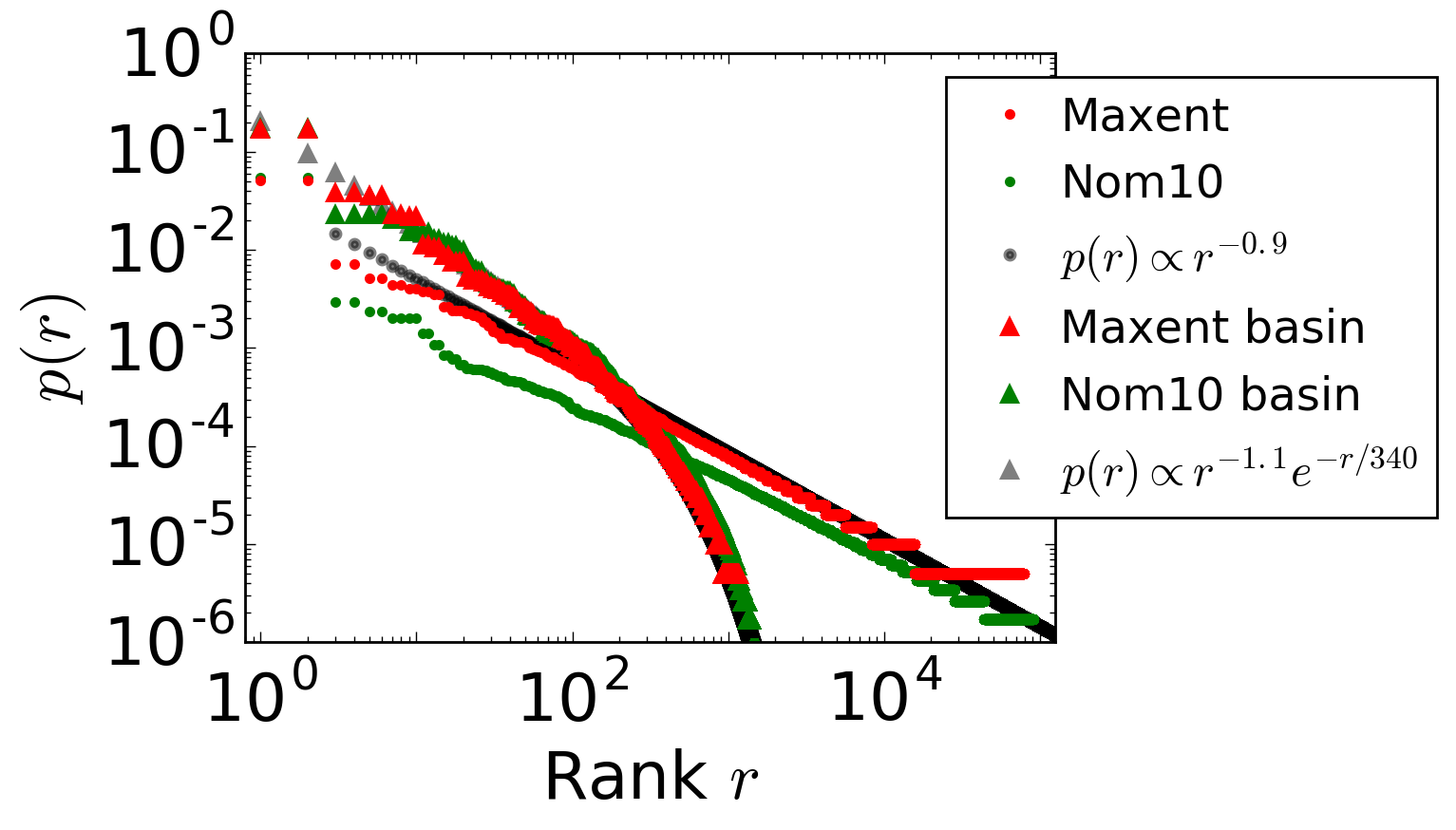}
\caption{(circles) Distribution of votes $\sigma$ obeys Zipf's law when ordered by rank $r$ as in Eq~\ref{eq:zipf}. Symmetry about the sign of the vote means that there are always two votes (one the negative of the other) with the same probabilities. The distribution is very heavy-tailed with exponent $\alpha=0.9$ according to the maxent model. (triangles) Distribution over energy basins follows a truncated power law (Eq~\ref{eq:zipf trunc}) with large cutoff \mbox{$\lambda = 340$}. Nom10 does not have a natural energy landscape relating votes, but shows remarkable overlap when assuming the maxent Hamiltonian in Eq~\ref{eq:E}.}
\label{gr:zipf}
\end{figure}

If I were, however, to group votes together such that all votes almost along ideological lines---where a few individuals buck the line---I might remove the heavy tail and find dominant clusters of ``noisy'' ideological votes that serve as a useful, compressed description of the Super Court \cite{Hopfield:1982vm,Schneidman:2006he,Lee:2015ev}. A natural way of clustering votes is given by the energy landscape constructed using Eq~\ref{eq:E}. By taking a vote and flipping the single voter most likely to change his or her vote at a time, I will eventually reach an energy minimum where no vote flip will lower the energy of the vote. This defines an energy landscape where states can be grouped by local energy minima. Since the energy is directly related to the probability, $\log p(\sigma) \propto -E(\sigma)$, this procedure is equivalent to searching for local maxima in probability where single voter flips climb hills in a probability landscape.
When I measure the probability that the Super Court is located on a particular hill of probability, I find again Zipf's law but now with an exponential cutoff 
\begin{align}
	p_{\rm basin}(r) \propto r^{-\alpha}e^{-r/\lambda} \label{eq:zipf trunc}
\end{align}
with $\alpha = 1.1$ and $\lambda= 340$. The exponential cutoff means that the clusters of noisy prototypical votes can indeed be summarized by dominant modes, but the size of the cutoff means the modes are many.

Although I cannot construct an energy landscape for W-Nominate in the same way because there are no explicit couplings between voters, if I were to assume that its energy landscape were structured similarly I should recover the same truncated Zipf's law using Eq~\ref{eq:E}. As I show in Figure~\ref{gr:zipf}, the resulting distribution over basins is remarkably similar. In comparison, random, independent voters show strong deviations from Eq~\ref{eq:zipf trunc} because they are randomly scattered across the entire extent of energy basins. 

The large number of energy minima is in stark contrast with the Second Rehnquist Court where there are only six meaningful clusters despite being a quarter of the size of the Super Court \cite{Lee:2015ev}. This comparison, however, is misleading. For spin glasses, systems with many frustrated interactions as with the Super Court, the number of maxima grows exponentially with system size \cite{Moore:1981iw,Bialek:2012ueb}, so the idea that the number of maxima scales linearly with size is strongly mistaken. This pattern is reflected in the eigenvalue spectrum of the covariance matrix $C_{\rm ij} = \br{\si{i}\si{j}}$ (Appendix Section IV) \cite{Sirovich:2003wt}. By calculating the spectrum on random subsets of N from the Super Court, I find the eigenvalues ordered by rank $r$ have the scale-free form
\begin{align}
	\lambda(r) &= N^\alpha \Lambda (rN^{-\alpha})
\end{align}
with $\alpha=0.8$. This means that there is no natural length scale at which to cut off the spectrum because every mode reveals another level of complexity in Super Court voting.

Dominant modes of voting are indeed useful and accurate descriptions of voting on natural courts because the systems are small \cite{Sirovich:2003wt,Lawson:2006hp}. Yet a low-dimensional representation of collective voting over time misses the heavy tail generating a rich range of behavior.

\section*{Discussion}
Although voting in a political context is complex, such collective phenomena are natural to approach from the perspective of statistical physics \cite{Schneidman:2006he,Guimera:2011ht,Daniels:2017cq,Bialek:uy,Bialek:2012ueb}.
Voting on the Supreme Court shows strong collective effects---often obscured by the focus on individual ideology \cite{Martin:2004vd,Epstein:2010wt,Segal:2015dn}---that I exploit by using observations of justices that voted together to infer how justices who never did might have (Figure~\ref{gr:sisj}). I find that the hypothetical Super Court is dominated by strong consensus that nearly spans a century (Figure~\ref{gr:unanimity}), an observation that may be surprising to Supreme Court scholars \cite{Walker:1988wh}. Hiding beneath this dominant mode is a rich range of dissenting coalitions (Figure~\ref{gr:zipf}), illustrating the hypothetical behavior of historically disjoint groups of justices. The remarkable similarity in the distribution of states beyond the unanimous mode of the maxent and W-Nominate models suggests that they are largely capturing the same collective structure encoded in the maxent energy landscape.

Consideration of such hypothetical situations has long played a significant role in legal research. Legal scholars study the influence of past precedents by using written opinions as sources of legal reasoning for justices in the future \cite{Urofsky:2017ug}. 
Political scientists compare justices by considering similarities and differences between the issues brought up in cases \cite{Baum:1989je}.
These alternative approaches for finding correlations across time are ways of indirectly testing the predictions of this model. 
A direct quantitative test might be for forecasting where the temporal structure captured by the pairwise maxent model might find immediate application in the prediction of Supreme Court votes \cite{Katz:2017gs}.

In contrast to binary ideological intuition obtained from news about the Supreme Court, the collective behavior of the Court over time reveals an institution focused on consensus and insulated from the seemingly rapid pace of political change. 
When I look beyond consensus, ideological intuition fails again if I account for the changing composition of the Court: ideological modes are lost amongst the cacophony of dissenting voices.
As we know from statistical physics, even simple models can generate an incredibly rich range of behavior when there are many competing, frustrated interactions \cite{Sherrington:1975ew,Nishimori:2001vha}. From the perspective of statistical physics, I show that this intuition is expressed in the complex voting patterns of the Supreme Court.

\begin{acknowledgments}
I thank Paul Ginsparg, Bryan Daniels, and Colin Clement for invaluable feedback and Bill Bialek, Guru Khalsa, and Ti-Yen Lan for comments on previous versions of the manuscript. I acknowledge funding from the NSF Graduate Research Fellowship under grant DGE-1650441.
\end{acknowledgments}

\bibliography{refs}

\begin{thebibliography}{10}

\bibitem{Brazill:2002te}
Bernard Grofman and Timothy~J Brazill.
\newblock {Identifying the median justice on the Supreme Court through
  multidimensional scaling: Analysis of {\textquotedblleft}natural
  courts{\textquotedblright} 1953{\textendash}1991}.
\newblock {\em Public Choice}, 112(1-2):55--79, 2002.

\bibitem{Martin:2004vd}
Andrew~D Martin, Kevin~M Quinn, and Lee Epstein.
\newblock {The Median Justice on the United States Supreme Court}.
\newblock {\em NCL Rev}, 83:1275, 2004.

\bibitem{Lawson:2006hp}
Brian~L Lawson, Michael~E Orrison, and David~T Uminsky.
\newblock {Spectral Analysis of the Supreme Court}.
\newblock {\em Math Mag}, 79(5):340, December 2006.

\bibitem{Urofsky:2017ug}
Melvin~I Urofsky.
\newblock {\em {Dissent and the Supreme Court}}.
\newblock Its Role in the Court's History and the Nation's Constitutional
  Dialogue. Vintage, January 2017.

\bibitem{Sirovich:2003wt}
Lawrence Sirovich.
\newblock {A pattern analysis of the second Rehnquist US Supreme Court}.
\newblock {\em PNAS}, 100(13):7432--7437, 2003.

\bibitem{Tenenbaum:2008tn}
Charles Kemp and Joshua~B Tenenbaum.
\newblock {The discovery of structural form}.
\newblock {\em PNAS}, 105(31):10687--10692, 2008.

\bibitem{Segal:2015dn}
Jeffrey~A Segal, Lee Epstein, Charles~M Cameron, and Harold~J Spaeth.
\newblock {Ideological Values and the Votes of U.S. Supreme Court Justices
  Revisited}.
\newblock {\em J of Politics}, October 2015.

\bibitem{Lee:2015ev}
Edward~D Lee, Chase~P Broedersz, and William Bialek.
\newblock {Statistical Mechanics of the US Supreme Court}.
\newblock {\em J Stat Phys}, pages 1--27, April 2015.

\bibitem{Tate:1981wy}
C~Neal Tate.
\newblock {Personal attribute models of the voting behavior of US Supreme Court
  justices: Liberalism in civil liberties and economics decisions,
  1946{\textendash}1978}.
\newblock {\em Am Political Sci Rev}, 75:355--367, 1981.

\bibitem{SupremeCourtD:YPR-mEko}
Harold~J Spaeth, Lee Epstein, Andrew~D Martin, Jeffrey~A Segal, Theodore~W
  Ruger, and Sara~C Benesh.
\newblock {2017 Supreme Court Database}.

\bibitem{States:TdVwx51P}
{Supreme Court of the United States}.
\newblock {The Supreme Court at Work}.

\bibitem{Quinn:2017ji}
Daniel~E Ho and Kevin~M Quinn.
\newblock {How Not to Lie with Judicial Votes: Misconceptions, Measurement, and
  Models}.
\newblock {\em California Law Review}, 98(3):813--876, 2010.

\bibitem{Poole:1985fo}
Keith~T Poole and Howard Rosenthal.
\newblock {A Spatial Model for Legislative Roll Call Analysis}.
\newblock {\em Am J of Political Sci}, 29(2):357, May 1985.

\bibitem{Shannon:1948wk}
Claude~Elwood Shannon.
\newblock {A Mathematical Theory of Communication}.
\newblock {\em The Bell System Technical Journal}, 27:379--423, July 1948.

\bibitem{Cover:2006tl}
Thomas~M Cover and Joy~A Thomas.
\newblock {\em {Elements of Information Theory}}.
\newblock John Wiley {\&} Sons, Hoboken, 2nd edition, 2006.

\bibitem{Jaynes:1957fy}
E~T Jaynes.
\newblock {Information Theory and Statistical Mechanics}.
\newblock {\em Phys Rev}, 106(4):620, May 1957.

\bibitem{Bialek:2012ueb}
W~S Bialek.
\newblock {\em {Biophysics: Searching for Principles}}.
\newblock Princeton University Press, 2012.

\bibitem{Epstein:2010wt}
Lee Epstein, Jeffrey~A Segal, and Harold~J Spaeth.
\newblock {The norm of consensus on the US Supreme Court}.
\newblock {\em Am J of Political Sci}, pages 362--377, May 2010.

\bibitem{Note1}
I use the probability of unanimity of a group of size $k'$ to estimate the
  probability that a single justice defects $p_{\protect \rm def} =
  1-p_{\protect \rm u}^{1/k'}(k')$. The probability of no defections with $k$
  independently defecting justices is $p_{\protect \rm u}(k) = (1-p_{\protect
  \rm def})^k$ as in Figure~\ref {gr:unanimity}.

\bibitem{Ising:1924vf}
E~Ising.
\newblock {\em {Beitrag zur Theorie des Ferromagnetismus}}.
\newblock PhD thesis, December 1924.

\bibitem{Newman:2005gv}
MEJ Newman.
\newblock {Power laws, Pareto distributions and Zipf's law}.
\newblock {\em Contemp Phys}, 46(5):323--351, September 2005.

\bibitem{Schwab:2014io}
David~J Schwab, Ilya Nemenman, and Pankaj Mehta.
\newblock {Zipf{\textquoteright}s Law and Criticality in Multivariate Data
  without Fine-Tuning}.
\newblock {\em Phys Rev Lett}, 113(6):068102, August 2014.

\bibitem{Hopfield:1982vm}
J~J Hopfield.
\newblock {Neural networks and physical systems with emergent collective
  computational abilities}.
\newblock {\em PNAS}, 79(8):2554--2558, April 1982.

\bibitem{Schneidman:2006he}
Elad Schneidman, Michael~J Berry~II, Ronen Segev, and William Bialek.
\newblock {Weak pairwise correlations imply strongly correlated network states
  in a neural population}.
\newblock {\em Nature}, 440(20):1007--1012, 2006.

\bibitem{Moore:1981iw}
A~J Bray and M~A Moore.
\newblock {Metastable states, internal field distributions and magnetic
  excitations in spin glasses}.
\newblock {\em J. Phys. C: Solid State Phys.}, 14(19):2629--2664, July 1981.

\bibitem{Guimera:2011ht}
Roger Guimer{\`a} and Marta Sales-Pardo.
\newblock {Justice Blocks and Predictability of U.S. Supreme Court Votes}.
\newblock {\em PLoS ONE}, 6(11):e27188, November 2011.

\bibitem{Daniels:2017cq}
Bryan~C Daniels, David~C Krakauer, and Jessica~C Flack.
\newblock {Control of finite critical behaviour in a small-scale social
  system}.
\newblock {\em Nat Comms}, 8:14301--8, February 2017.

\bibitem{Bialek:uy}
W~Bialek, A~Cavagna, and I~Giardina.
\newblock {Social interactions dominate speed control in poising natural flocks
  near criticality}.
\newblock {\em PNAS}, 2014.

\bibitem{Walker:1988wh}
T~G Walker, L~Epstein, and WJ~Dixon.
\newblock {On the mysterious demise of consensual norms in the United States
  Supreme Court}.
\newblock {\em J of Politics}, 50(2):361--389, 1988.

\bibitem{Baum:1989je}
L~Baum.
\newblock {Comparing the Policy Positions of Supreme Court Justices from
  Different Periods}.
\newblock {\em Political Res Q}, 42(4):509--521, December 1989.

\bibitem{Katz:2017gs}
Daniel~Martin Katz, Michael~J Bommarito, and Josh Blackman.
\newblock {A general approach for predicting the behavior of the Supreme Court
  of the United States}.
\newblock {\em PLoS ONE}, 12(4):e0174698--18, April 2017.

\bibitem{Sherrington:1975ew}
David Sherrington and Scott Kirkpatrick.
\newblock {Solvable Model of a Spin-Glass}.
\newblock {\em Phys Rev Lett}, 35(26):1792--1796, December 1975.

\bibitem{Nishimori:2001vha}
Hidetoshi Nishimori.
\newblock {\em {Statistical Physics of Spin Glasses and Information
  Processing}}.
\newblock An Introduction. Clarendon Press, 2001.

\end{thebibliography}

\end{document}